\documentclass[12pt]{iopart}

\usepackage{iopams}
\usepackage{graphicx}
\usepackage{subfig}
\usepackage{color}
\usepackage{ulem}
\usepackage{color}
\usepackage{ulem}

\begin{document}

\title{MHD simulations of cold bubble formation from 2/1 tearing mode during massive gas injection in a tokamak}

\author{Shiyong Zeng}
\address{Department of plasma physics and fusion engineering, University of Science and Technology of China, Hefei, Anhui 230026, China}

\author{Ping Zhu}
\address{International Joint Research Laboratory of Magnetic Confinement Fusion and Plasma Physics, State Key Laboratory of Advanced Electromagnetic Engineering and Technology, School of Electrical and Electronic Engineering, Huazhong University of Science and Technology, Wuhan, Hubei 430074, China}
\address{Department of Engineering Physics, University of Wisconsin-Madison, Madison, Wisconsin 53706, USA}
\ead{zhup@hust.edu.cn}

\author{V.A. Izzo}
\address{Fiat Lux, San Diego, CA, United States of America}

\author{Haolong Li}
\address{College of Physics and Optoelectronic Engineering, Shenzhen University, Shenzhen 518060, China}

\author{Zhonghe Jiang}
\address{International Joint Research Laboratory of Magnetic Confinement Fusion and Plasma Physics, State Key Laboratory of Advanced Electromagnetic Engineering and Technology, School of Electrical and Electronic Engineering, Huazhong University of Science and Technology, Wuhan, Hubei 430074, China}

\title[Simulations of cold bubble formation from 2/1 tearing mode during MGI]{}

\newpage
\begin{abstract}
Massive gas injection (MGI) experiments have been carried out in many tokamaks to study disruption dynamics and mitigation schemes. Two events often observed in those experiments are the excitation of the $m=2, n=1$ magnetohydrodynamic (MHD) mode, and the formation of cold bubble structure in the temperature distribution before the thermal quench (TQ). Here $m$ is the poloidal mode number, $n$ the toroidal mode number. The physics mechanisms underlying those phenomena, however, have not been entirely clear.
In this work, our recent NIMROD simulations of the MGI process in a tokamak have reproduced the main features of both events, which has allowed us to examine and establish the causal relation between them. In these simulations, the $3/1$ and $2/1$ islands are found to form successively after the arrival of impurity ion cold front at the corresponding $q = 3$ and $q = 2$ rational surfaces. At the interface between impurity and plasma, a local thin current sheet forms due to an enhanced local pressure gradient and moves inward following the gas cold front, this may contribute to the formation of a dominant $2/1$ mode.
Following the growth of the $2/1$ tearing mode, the impurity penetration into the core region inside the $q=2$ surface gives rise to the formation of the cold bubble temperature structure and initiates the final TQ. A subdominant $1/1$ mode developed earlier near the $q=1$ surface alone does not cause such a cold bubble formation, however, the exact manner of the preceding impurity penetration depends on the nature of the $1/1$ mode: kink-tearing or quasi-interchange.
\end{abstract}

\section{Introduction}
\label{sec1}


Macroscopic instabilities in tokamaks can largely degrade plasma performance, cause abrupt discharge termination and severely threaten steady operation of devices. Without proper mitigation, disruptions can deposit substantial heat loads, unbalanced electromagnetic forces, and runaway electron current to the first wall and plasma facing components, causing disastrous damage to the machine \cite{Hender2007}. Disruption mitigation schemes based on the massive gas injection (MGI) method have been widely studied on major tokamaks including JET \cite{Lehnen2015a,wesson1989}, D\uppercase\expandafter{\romannumeral3}-D \cite{Hollmann2007,Shiraki2015,Eidietis2017,Shiraki2016}, ASDEX-Upgrade \cite{AsdexUpgradeTeam2016,Pautasso2017}, KSTAR \cite{Choi2016}, EAST \cite{Chen2018}, J-TEXT \cite{Tong2019,Tong2018a,Huang2018,DING_2018}. Although recent designs for the ITER disruption mitigation scheme have opted toward the more efficient shattered pellet injection (SPI) system, the MGI system has remained viable and effective for disruption mitigation on most tokamaks, at least during the thermal quench (TQ) phase \cite{Hollmann2015}. Meanwhile, simulations of MGI have been performed using NIMROD \cite{Izzo2008,Izzo2013,Izzo2015,Izzo2017}, JOREK \cite{Fil2015}, and M3D-C1 \cite{Ferraro_2018} codes, and comprehensive and systematical comparison has been performed between the codes and experiments. For example, NIMROD simulations reproduce the sequence of events observed in MGI experiments and demonstrate the relationship between locked modes and the thermal quench \cite{Izzo2008,Sweeney_2018}, JOREK simulations show the island formation and mode growth during the MGI process \cite{Nardon2017}.
Despite this progresses, some key phenomena observed during MGI experiments have not been well understood. Among them, the causal relation, if any, between onset of the $m=2, n=1$ tearing mode and the formation of cold bubble has remained unclear. Here $m$ ($n$) is the poloidal (toroidal) mode number.


MGI experiments often observe the $2/1$ MHD mode that dominates the mitigation process and leads to the TQ.
Most MGI experiments also find impurity penetration shallow, which typically stops outside the vicinity of the $q = 2$ surface. For example, in Tore Supra experiments, bursts of MHD instability occur after the gas cold front stops along the $q = 2$ surface \cite{Reux2010}.
In J-TEXT experiments, impurity penetration and assimilation are enhanced when the $2/1$ mode width grows above a critical value, which accelerates the thermal quench process \cite{Tong2019}.
In addition, the locked or quasi-stationary modes, usually the $m = 2, n = 1$, have been found before disruption in many devices such as D\uppercase\expandafter{\romannumeral3}-D \cite{Sweeney2017}.


With respect to how the $2/1$ mode leads to the TQ, D\uppercase\expandafter{\romannumeral3}-D experiments show that the closer the $q = 2$ surface is located towards the separatrix, the sooner the TQ may launch \cite{Hollmann2007}, which suggests the correlation between the $2/1$ mode and the onset of TQ. Simulations of density limit disruption indicate that the $2/1$ mode can couple with the $1/1$ mode, which may involve connection through the $3/2$ mode. The coupling eventually leads to the explosive growth of the $ m \geq 2,n=1 $ modes and the complete stochasticity along with the current profile broadening \cite{Bondeson1991}.
Besides, some simulation results propose that the $2/1$ magnetic island can grow to fill a substantial part of the poloidal plane to trigger the major disruption \cite{White1997,Waddell1978,Sykes1980}.

Another universally observed phenomenon in both MGI and density limit disruption experiments is the formation of $1/1$ temperature structure, also known as ``cold bubble", observed from SRX signal during the final disruption phase \cite{Howard1992}.
In KSTAR experiments, the cold bubble can grow from and couple with the $2/1$ island to give rise to major disruption \cite{Choi2016}.
MGI experiments on JET show that it is from the reconnection region (X-point) that the hot core plasma is expelled, and the O-point is where the colder plasma outside is absorbed \cite{Lehnen2015a}. Similar results are found in J-TEXT experiments as well \cite{Tong2019}.
Gates \cite{Gates2012} proposed that the cold bubble is caused by a $1/1$ radiation driven island based on their theory model for Greenwald density limit. Actually the $1/1$ mode temperature structure itself is also found in previous NIMROD simulations \cite{Izzo2013}, however, that study focused on other aspects of the MGI process, for example, impurity assimilation efficiency and radiation asymmetry.

In this work, the MGI disruption mitigation process in a tokamak is simulated using the $3D$ extended MHD code NIMROD \cite{Izzo2008,Izzo2013}, which incorporates an atomic and radiation physics model from KPRAD \cite{KPRAD}.
The purpose of this work is to understand the physical connection between the two often observed phenomena before thermal quench during the MGI process, namely the onset of 2/1 tearing mode and the formation of cold bubble. Such understanding may also help us to explore the physics underlying the similar process in density limit disruptions.
Our analysis of the NIMROD simulation results may explain how the $2/1$ tearing mode may contribute to the formation of cold bubble and the start of TQ. A dominant $2/1$ mode is found due to the gas cold front penetration. In particular, at the location in the poloidal plane where the impurity ion cold front is aligned with the X-point or O-point of the $2/1$ mode, the impurity gas penetrates further into the core, giving rise to the formation of cold bubble and the start of TQ.

The rest of the paper is organized as follows. Section 2 describes the simulation model and setup. Section 3 shows the overall simulation results on the MGI process as a function of time. Section 4 focuses on the island growth on rational surfaces, the onset of $2/1$ modes, and the formation of cold bubble during the MGI process. Section 5 gives a discussion and summary.


\section{NIMROD/KPRAD model and simulation setup}
\label{sec2}
Our simulations in this work are based on the single-fluid resistive MHD model implemented in the NIMROD code \cite{SOVINEC2004355}, and a simplified module for impurity radiation adapted from the KPRAD code. The equations for the impurity-MHD model are as follows:

\begin{eqnarray}
\rho \frac{d\vec{V}}{dt} = - \nabla p + \vec{J} \times \vec{B} + \nabla \cdot (\rho \nu \nabla \vec{V})
\label{eq:momentum}
\\
\frac{d n_i}{dt} + n_i \nabla \cdot \vec{V} = \nabla \cdot (D \nabla n_i) + S_{ion/3-body}
\label{eq:contiune2}
\\
\frac{d n_{Z,Z=0-18}}{dt} + n_Z \nabla \cdot \vec{V} = \nabla \cdot (D \nabla n_Z) + S_{ion/rec}
\label{eq:contiune3}
\\
n_e \frac{d T_e}{dt} = (\gamma - 1)[n_e T_e \nabla \cdot \vec{V} + \nabla \cdot \vec{q_e} - Q_{loss}]
\label{eq:temperature}
\\
\vec{q}_e = -n_e[\kappa_{\parallel} \hat{b} \hat{b} + \kappa_{\perp} (\mathcal{I} - \hat{b} \hat{b})] \cdot \nabla T_e
\label{eq:heat_flux}
\\
\vec{E} + \vec{V} \times \vec{B} = \eta \vec{j}
\label{eq:ohm}
\end{eqnarray}

Here, $n_i$, $n_e$, and $n_Z$ are the main ion, electron, and impurity ion number density respectively, $\rho$, $\vec{V}$, $\vec{J}$, and $p$ the plasma mass density, velocity, current density, and pressure respectively. $T_e$ and $\vec{q}_e$ the electron temperature and heat flux respectively. $D$, $\nu$, $\eta$, and $\kappa_{\parallel} (\kappa_{\perp})$ the plasma diffusivity, kinematic viscosity, resistivity, and parallel (perpendicular) thermal conductivity respectively, $\gamma$ the adiabatic index, $S_{ion/rec}$ the density source from ionization and recombination, $S_{ion/3-body}$ also includes contribution from 3-body recombination, $Q_{loss}$ the energy loss, $\vec{E} (\vec{B})$ the electric (magnetic) field, $\hat{b}=\vec{B}/B$, and $\mathcal{I}$ the unit dyadic tensor.

All particle species share a single temperature $T=T_e$ and fluid velocity $V$, which assumes instant thermal equilibration between main ion and impurity species. Pressure $p$ and mass density $\rho$ in momentum equation (\ref{eq:momentum}) include impurity contributions.
Each charge state of impurity ion density is tracked in the KPRAD module and used to update the source/sink terms in the continuity equations due to ionization and recombination \cite{Izzo2008}. Both convection and diffusion terms are included in each continuity equation where all the diffusivities are the same.
Quasi-neutrality is maintained through $n_e=n_i+\sum Z n_{z}$, where $Z$ is the charge of impurity ion.
The energy source term $Q_{loss}$ in equation (\ref{eq:temperature}) is calculated from the KPRAD module based on a coronal non-equilibrium model, which includes energy loss from bremsstrahlung, line radiation, ionization, recombination, and background impurity radiation \cite{KPRAD}. The energy gain from ohmic heating is then added to the source term.
Anisotropic thermal conductivities are temperature dependent, i.e. $\kappa_{\parallel} \propto T^{5/2}$ and $\kappa_{\perp} \propto T^{-1/2}$.
Finally, the temperature-dependence in the Spizter model for resistivity $\eta$ is believed to be a key physics factor for the accurate simulation of the TQ \cite{doi:10.1063/1.5088814}.

\begin{figure}[ht]
  \begin{center}
  \includegraphics[width=0.45\textwidth,height=0.3\textheight]{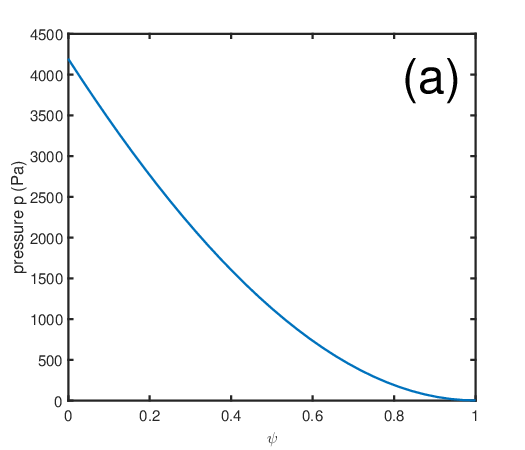}
  \includegraphics[width=0.45\textwidth,height=0.3\textheight]{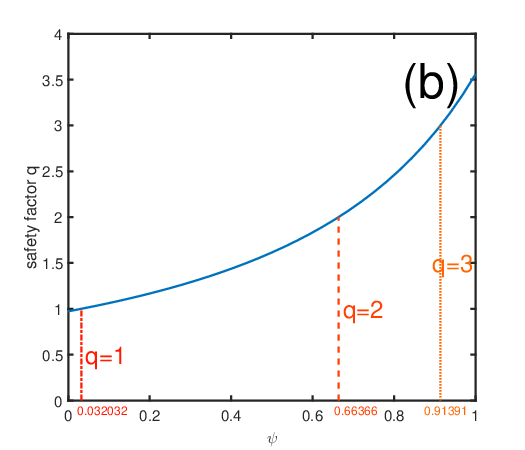}
  \includegraphics[width=0.5\textwidth,height=0.3\textheight]{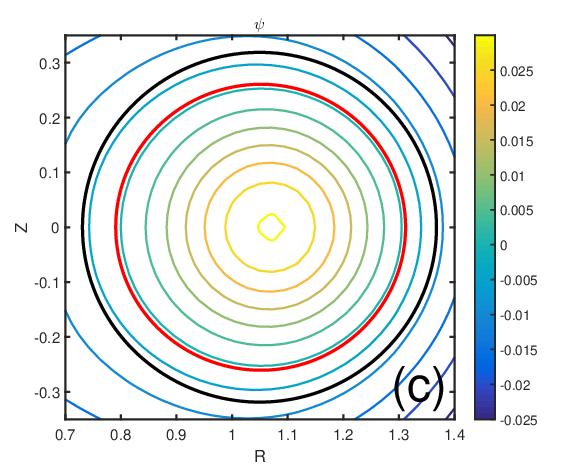}
  \end{center}
  \caption{(a) Pressure $p$, (b) safety factor $q$ as functions of the normalized flux function $\psi$ for the J-TEXT like equilibrium obtained from EFIT calculation and used in this work. $q = 1,2,3$ surfaces are denoted as vertical broken lines, (c) the initial equilibrium magnetic flux $\psi$, where the simulation domain boundary is denoted as the black circle and the plasma boundary is denoted as the red circle.}
\label{equilibrium}
\end{figure}

For simplicity, a J-TEXT like tokamak equilibrium with a circular shaped boundary is considered in this work (Fig. \ref{equilibrium}). Key equilibrium and input parameters are listed in table \ref{input}.
The initial impurity distribution is localized right outside plasma boundary, which assumes the following form
\begin{equation}
 S_{imp}=n_{imp} \left [ 100 \tanh{ \left ( \frac{r}{r_v}-1 \right ) } +1 \right ]  \exp{ \left [ -\left ( \frac{\theta - \theta_0}{15} \right ) ^2  -\left ( \frac{\phi - \phi_0}{15} \right ) ^2 \right ] }.
 \label{imp}
\end{equation}
Here $n_{imp}$ is the injected impurity density, $r_v$ the radius of plasma boundary, $\theta_0$ ($\phi_0$) the poloidal (toroidal) angle of the impurity gas injection location.
We use $64 \times 63$ grids and third order polynomial of Lagrange-type finite elements in the poloidal plane, a semi-implicit time-advance is applied. The plasma is limited by a perfect conducting wall, and the boundary of the simulation domain is surrounded by a vacuum region.

\begin{table}
\caption{\label{input} Key parameters in the simulation}
\footnotesize
\begin{tabular}{@{}llll}
\br
Parameter & Symbol & Value & Unit \\
\mr
Minor radius & $a$ & $0.25$ & $m$ \\
Major radius & $R_0$ & $1.05$ & $m$ \\
Plasma current & $I_p$ & $150$ & kA \\
Toroidal magnetic field & $B_{t0}$ & $1.75$ & T\\
Edge value of safety factor & $q_a$ & $3.56$ & dimensionless \\
Core electron number density & $n_{e0}$ & $1.875 \times 10^{19}$ & $m^{-3}$ \\
Core electron temperature & $T_{e0}$ & $700$ & $eV$\\
Equilibrium velocity & $V_0$ & $0$ & $m/s$\\
The core Lundquist number & $S_0$ & $4.055 \times 10^7$  & dimensionless \\
Core perpendicular thermal conductivity & $\kappa_{\perp0}$ & $1$ & $m^2/s$ \\
Core parallel thermal conductivity & $\kappa_{\parallel0}$ & $10^6$ & $m^2/s$ \\
Diffusivity & $D$ & $2$ & $m^2/s$\\
\br
\end{tabular}\\
\end{table}

\section{Time history of MGI process from NIMROD simulation}
\begin{figure}[ht]
  \begin{center}
  \includegraphics[width=0.8\textwidth,height=0.8\textheight]{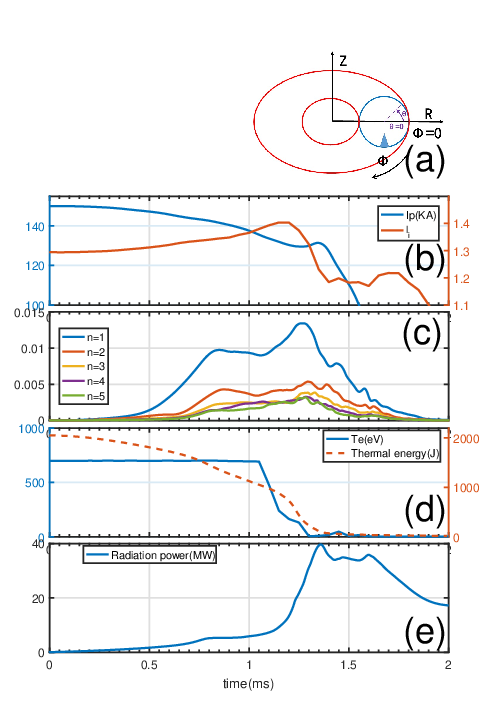}
  \end{center}
  \caption{(a) Sketch of coordinate system showing the initial impurity injection from the blue triangle region ($\phi_0 = 0,\theta_0 = 270$). (b) Plasma current (blue solid line) and internal inductance (red solid line), (c) normalized magnetic energies of toroidal components $\sqrt(W_{mag,n}/W_{mag,n=0})$, (d) core electron temperature (blue solid line) and thermal energy (red dashed line), and (e) radiation power as functions of time during a MGI process from NIMROD simulation, where $0-1.05 ms$ is the pre-TQ phase, and $1.05-1.35 ms$ is the TQ phase.}
\label{discharge}
\end{figure}

Our NIMROD simulations have reproduced the main features of the MGI process often observed in experiments. For an impurity Ar gas initially injected from the plasma boundary at the angle of ($\phi_0=0$, $\theta_0=270$) i.e. the bottom of a poloidal plane, the pre-thermal quench (pre-TQ) is identified as the period from $0$ to $1.05 ms$, which is characterized with a gradual decay (increase) in thermal energy (radiation power) (Figs. \ref{discharge}d-\ref{discharge}e). During the pre-TQ phase, the $n = 1-5$ MHD modes start to grow after $t = 0.5 ms$ and saturate at $t = 0.8 ms$, where the $n = 1$ mode dominates the growth (Fig. \ref{discharge}c).

The TQ phase starts with a sudden sharp drop in the core electron temperature at $t = 1.05 ms$, and ends with a current spike at $t = 1.35 ms$. During the TQ phase, all magnetic surfaces in the core region are completely destroyed and the current profile broadens. Subsequently, the plasma totally cools down and loses confinement, the current profile expands outwards which is identified by the decrease of the internal inductance $l_i$ and the appearance of the current spike. The $n = 1$ mode continues to grow at the same time of the core electron temperature drop and its amplitude reaches maximum later. The radiation power surges after the collapse of temperature, and reaches a peak by the end of TQ phase (Fig. \ref{discharge}e).
The current quench (CQ) phase follows immediately afterward, during which the radiation power remains large and balanced with the Ohmic heating power due to the enhanced resistivity and slowly decaying plasma current. The CQ phase is not the focus of this study, however.

\section{Onset of 2/1 tearing modes and formation of cold bubble}

\subsection{Impurity penetration and island growth at rational surfaces}
\begin{figure}[ht]
  \begin{center}
  \includegraphics[width=0.45\textwidth,height=0.35\textheight]{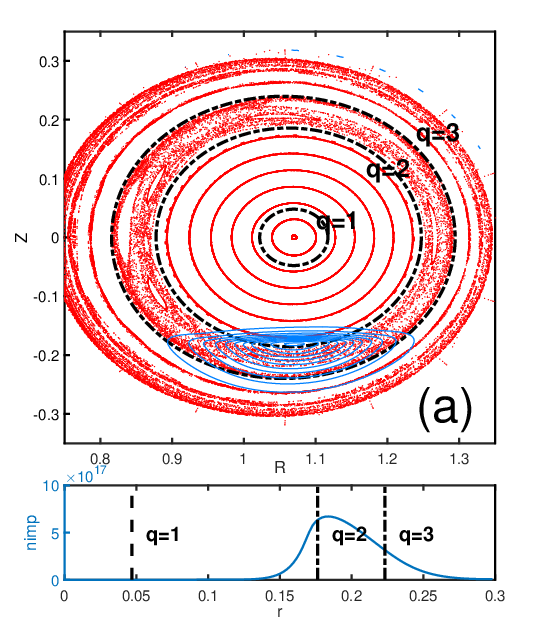}
  \includegraphics[width=0.45\textwidth,height=0.35\textheight]{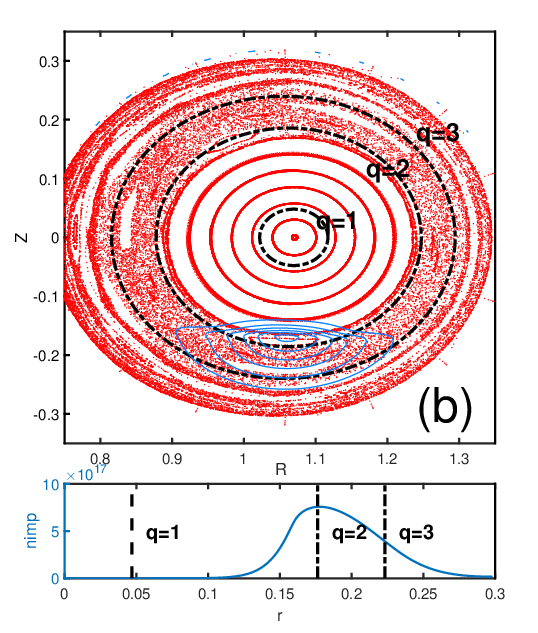}
  \includegraphics[width=0.45\textwidth,height=0.35\textheight]{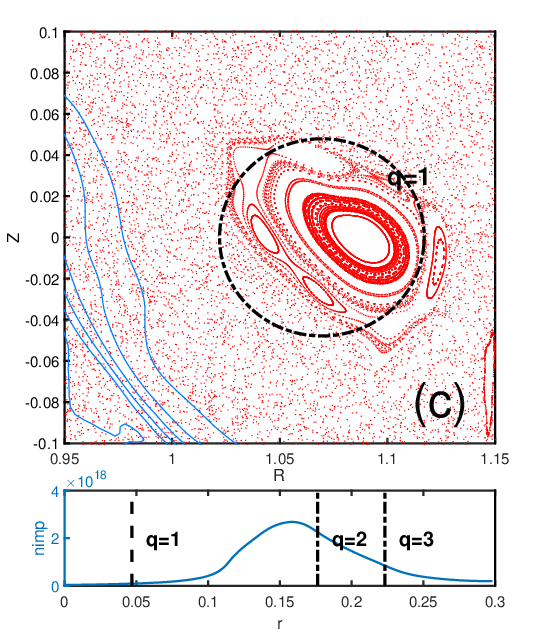}
  \end{center}
  \caption{Poincare plot (red dot) and impurity ion distribution (blue line, the sum of all Ar ion charge states) in the poloidal plane at toroidal angle $\phi=0$ (upper panel) and radial profile of the flux surface averaged impurity ion density (lower panel) at (a) $t = 0.25 ms$, (b) $t = 0.35 ms$, and (c) $t = 0.95 ms$. $q = 1,2,3$ surfaces are denoted as black dashed-line circles.}
\label{poincare}
\end{figure}

During the early stage after impurity injection, the $3/1$ island appears first after the arrival of the peak impurity ion density on the $q = 3$ surface from the boundary. After the peak impurity ion distribution reaches the $q = 2$ surface and accumulates there afterward (Fig .\ref{poincare}a), the $2/1$ mode is excited and dominates until well into the TQ phase (Fig. \ref{poincare}b). The gas cold front eventually penetrates inside the $q = 1$ surface when the last unbroken magnetic flux surface in the core region disappears after $t = 0.95 ms$, which initiates the TQ. Right before that, several smaller secondary islands can be found in the vicinity of the $q = 1$ surface (Fig. \ref{poincare}c), which are related to the high n modes shown in Fig. \ref{discharge}(c). In addition, MGI experiments in J-TEXT have observed that the similar high $(m, n)$ modes, such as $3/2, 4/3, 5/4$ … start to grow right before TQ \cite{Tong2019}.

\begin{figure}[ht]
	\begin{center}
		\includegraphics[width=0.85\textwidth,height=0.55\textheight]{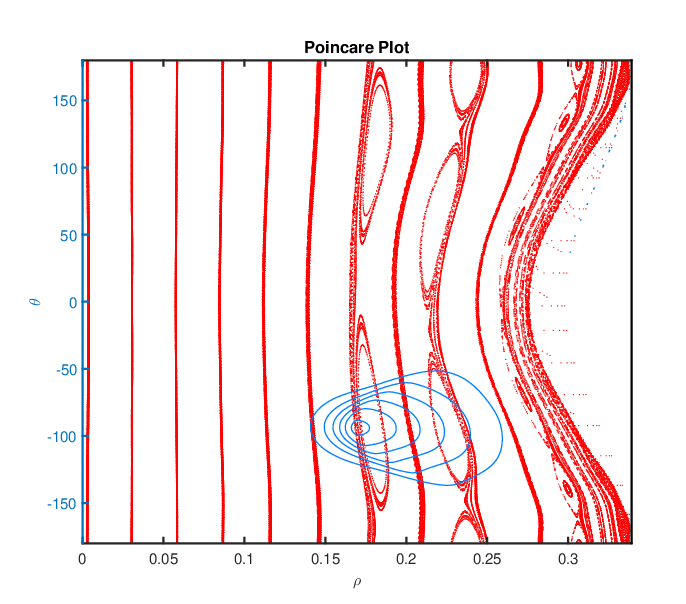}
	\end{center}
	\caption{Poincare plots of the magnetic fields including both the equilibrium and the $n=1$ components (red dot) and impurity ion distribution (blue line, the sum of all Ar ion charge states) in the $\rho,\theta$ poloidal plane at toroidal angle $\phi=0$, when $t = 0.35 ms$.}
	\label{poincare_plane}
\end{figure}

Whereas the impurity ion penetrates radially inward through diffusion and convection within the poloidal plane over time, the location of its cold front corresponds to the O-points of the $3/1$ and $2/1$ island (Fig. \ref{poincare_plane}). Similar phase alignment of those modes has also been identified in JOREK simulation results \cite{Nardon2017}.
According to the continuity equation, the impurity spreading is directly governed only by flows and density gradients, but the magnetic topology indirectly affects the impurity spreading in several important ways. First, it has been show that the impurities will spread more rapidly in the parallel direction on islands or rational surfaces than on irrational flux surfaces \cite{Izzo2015}, which consequently reduces the radial gradient and impedes inward spreading. Further, the parallel spreading will be dominantly toward the HFS due to the magnetic nozzle effect \cite{Izzo2015}. Additionally, in the simulation (despite zero equilibrium flow) the islands rotate clockwise in the poloidal plane, and the induced flows in the simulation can transport impurities both across and along field lines.

\subsection{Current sheet formation at the impurity-plasma interface}
\begin{figure}[ht]
  \begin{center}
  \includegraphics[width=0.49\textwidth,height=0.3\textheight]{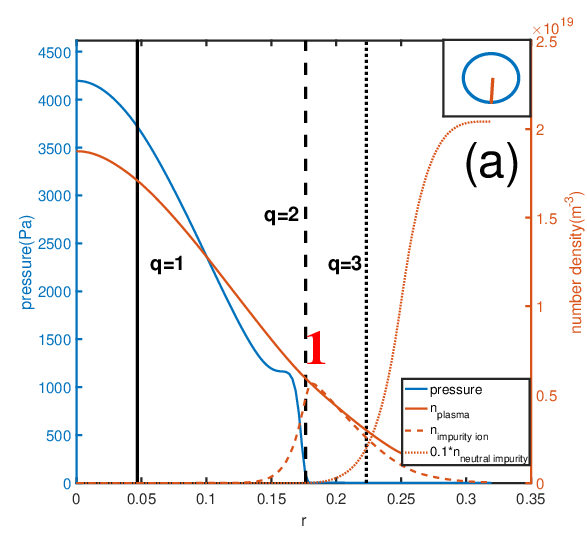}
  \includegraphics[width=0.49\textwidth,height=0.3\textheight]{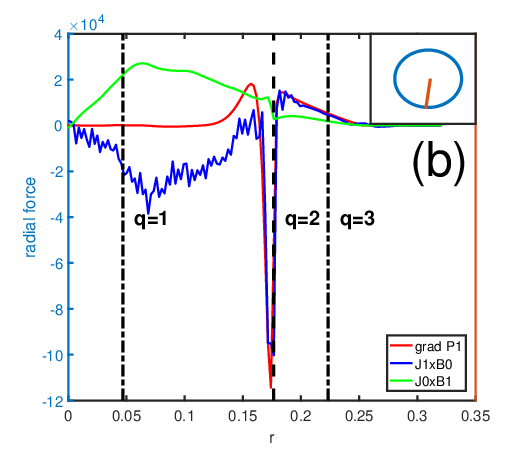}
  \includegraphics[width=0.49\textwidth,height=0.3\textheight]{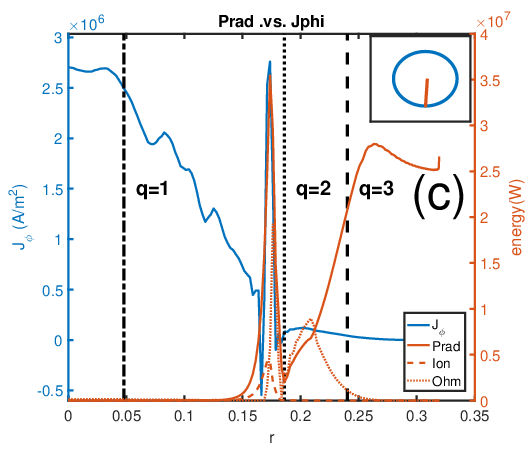}
  \end{center}
  \caption{Radial profiles along the $\theta=270$ line in the poloidal plane (red line in the embedded sketch) at toroidal angle $\phi=0$, $t = 0.25 ms$ for (a) pressure (blue solid curve), plasma density (red solid curve), impurity ion density (red dashed curve, the sum of all Ar ion charge states), and neutral impurity density (red dotted curve) (b) pressure gradient perturbation $\nabla p_1=\nabla(p-p_0)$ (red solid curve), and Lorentz force perturbations $J_1 \times B_0$ (blue solid curve) and $J_0 \times B_1$ (green solid curve) ($J_1 = J - J_0, B_1 = B - B_0$), and (c) toroidal plasma current density (blue solid curve), radiation power (red solid line), ionization power (red dashed line) and ohmic heating power (red dotted line). $q = 1,2,3$ surfaces are denoted as black vertical broken lines.}
\label{interface}
\end{figure}

In the poloidal plane of the toroidal injection angle $\phi_0=0$, the impurity ion cold front arrives at the $q = 2$ surface when $t = 0.25 ms$. We denote the location ``1" as the impurity-plasma interface where the impurity ion cold front has the same density level as the background plasma (Fig. \ref{interface}a). Inside the interface, the plasma is slightly perturbed and the magnetic flux surfaces remain intact. Outside the interface, where the bulk of neutral impurity are located, the plasma is nearly cooled down and the magnetic field lines become stochastic (Fig. \ref{poincare}b). The pressure profile is slightly flattened inside the interface, and the gradient at the interface becomes steeper than inside due to the radiative cooling from the impurities (Fig. \ref{interface}a).

A new radial force balance from $\nabla p = \vec{J} \times \vec{B}$ is established at the interface between the enhanced pressure gradient and the local Lorentz force, as indicated from Fig. \ref{interface}(b). Most importantly, through the new radial force balance, the enhanced radial pressure gradient leads to an enhanced local toroidal current density, i.e. the formation of a current sheet at the impurity-plasma interface near the $q = 2$ surface (Fig. \ref{interface}c).
Such a current sheet is accompanied by a sharp peak in the radiation power as well as the ionization profile, which leads to an enhanced ohmic heating power in the cold plasma region (Fig. \ref{interface}c).
The formation of this current sheet reinforces the equilibrium current density gradient at the $q = 2$ surface, thus may contribute to the onset of the $2/1$ modes. The current sheet is similar to the skin current formed in the previous M3D-C1 simulation \cite{Ferraro_2018}. Density limit disruption simulation has also found similar edge-cooling-induced current sheet formation that destabilizes a sequence of precursor modes ($2/1,3/2,...$) \cite{Kleva1991}.

\subsection{Current density contraction and the $2/1$ tearing mode}

\begin{figure}[ht]
  \begin{center}
  \includegraphics[width=0.64\textwidth,height=0.42\textheight]{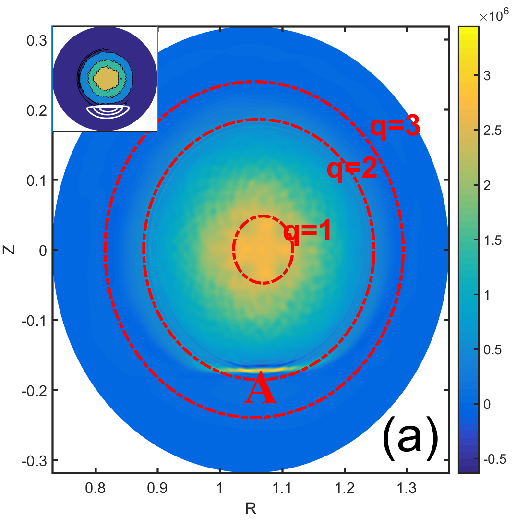}
  \includegraphics[width=0.64\textwidth,height=0.42\textheight]{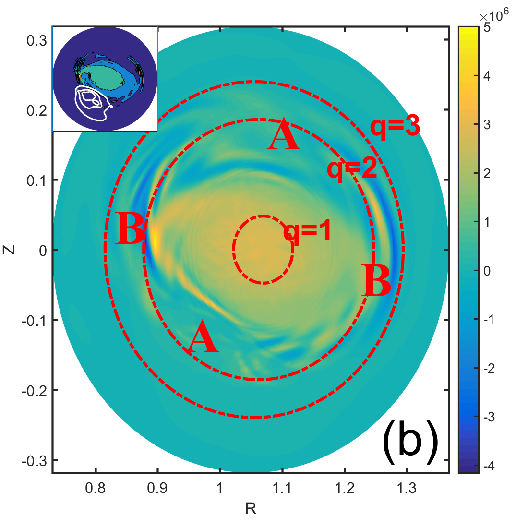}
  \end{center}
  \caption{Plasma current density distribution (in unit $A/m^2$, flushed color), and impurity ion distribution (white lines in the sketch) in the poloidal plane at toroidal angle $\phi =0 $ at (a) $t = 0.25 ms$, and (b) $t = 0.75 ms$. $q = 1,2,3$ surfaces are denoted as red line circles.}
\label{current_contour}
\end{figure}

Radiation cooling leads to the contraction of current density at the $q = 2$ surface upon its initial direct contact at $t = 0.25 ms$ with the impurities injected from the bottom of the poloidal plane at the region A shown in Fig. \ref{current_contour}(a). Whereas the magnetic surfaces inside the interface remain intact, the opposite top side of the current density distribution contracts subsequently due to fast parallel thermal transport as well. The maximum current density contraction is located at the O-point of magnetic island (Fig. \ref{current_contour}b region A). Then the entire current density distribution contracts with the impurity ion cold front penetration over time. In addition, the total plasma current barely changes during the pre-TQ phase, therefore the vertical contraction of current density results in the excess of current density at the two horizontal sides shown in region B of Fig. \ref{current_contour}(b). This gradually leads to the elliptical distribution of current density and the local current sheet formation at regions A and B within the poloidal plane (Fig. \ref{current_contour}b).

\begin{figure}[ht]
  \begin{center}
  \includegraphics[width=0.49\textwidth,height=0.45\textheight]{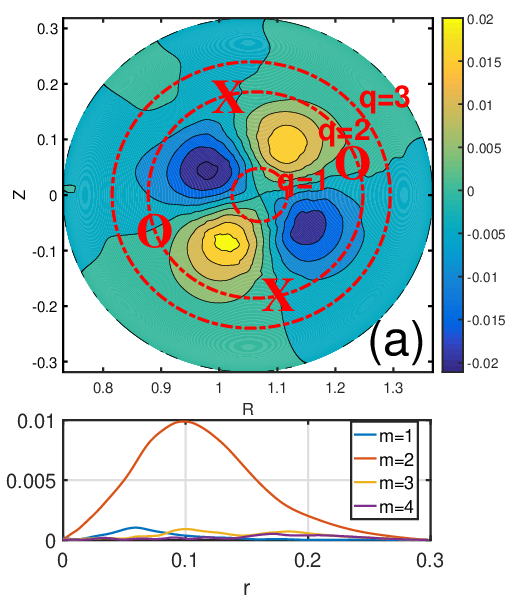}
  \includegraphics[width=0.49\textwidth,height=0.45\textheight]{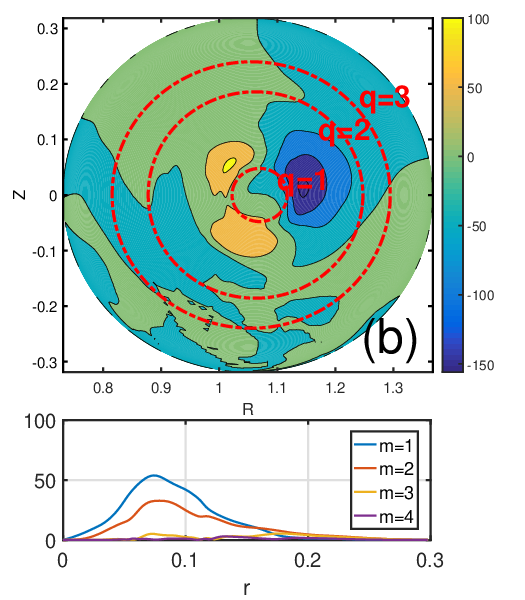}
  \end{center}
  \caption{(a) $n = 1$ mode structure of normal component of perturbed magnetic field $B_r$ contour (in unit $T$, upper panel) and radial profile of its poloidal Fourier component ($m$ refers to poloidal mode number, lower panel), (b)  $n = 1$ mode structure of electron temperature contour (in unit $eV$, upper panel) and radial profile of its poloidal Fourier component ($m$ refers to poloidal mode number, lower panel), both at toroidal angle $\phi =0 $ and $t = 1.05 ms$. $q = 1,2,3$ surfaces are denoted as red line circles.}
\label{FT}
\end{figure}

A dominant $2/1$ tearing mode can be found to peak in the region between the equilibrium $q = 2$ and $q = 1$ surfaces as a result of the current density contraction following the gas cold front penetration (Fig. \ref{FT}a).
The radial profile of the poloidal Fourier component of $B_r$ confirms its $2/1$ mode structure as well. Besides, the peak of the $m=2$ component profile is located inward of the equilibrium $q = 2$ surface, as a consequence of the current density contraction. The two peaking locations of the local current density accumulation (region B, Fig. \ref{current_contour}b) are consistent with the X-point locations of the dominant $2/1$ mode (Fig. \ref{FT}a), where the slight difference in poloidal angle is due to the clockwise rotation of island. It is worth noting that the maximum impurity ion density is located at the toroidal $\phi=0$ plane of initial injection before the TQ. Thus the interaction between the impurity and plasma takes place mainly within the toroidal $\phi=0$ plane, which may determine the phases of the subsequent tearing modes. Whereas the magnetic perturbation is dominated by the $2/1$ mode, the dominant mode component of the perturbed temperature has become $1/1$ right before the onset of TQ (Fig. \ref{FT}b), which used to be $2/1$ during the pre-TQ phase. This $1/1$ mode structure of temperature distribution is the so-called ``cold bubble".

\subsection{Cold bubble formation}
\begin{figure}[ht]
	\begin{center}
		\includegraphics[width=1.0\textwidth,height=0.3\textheight]{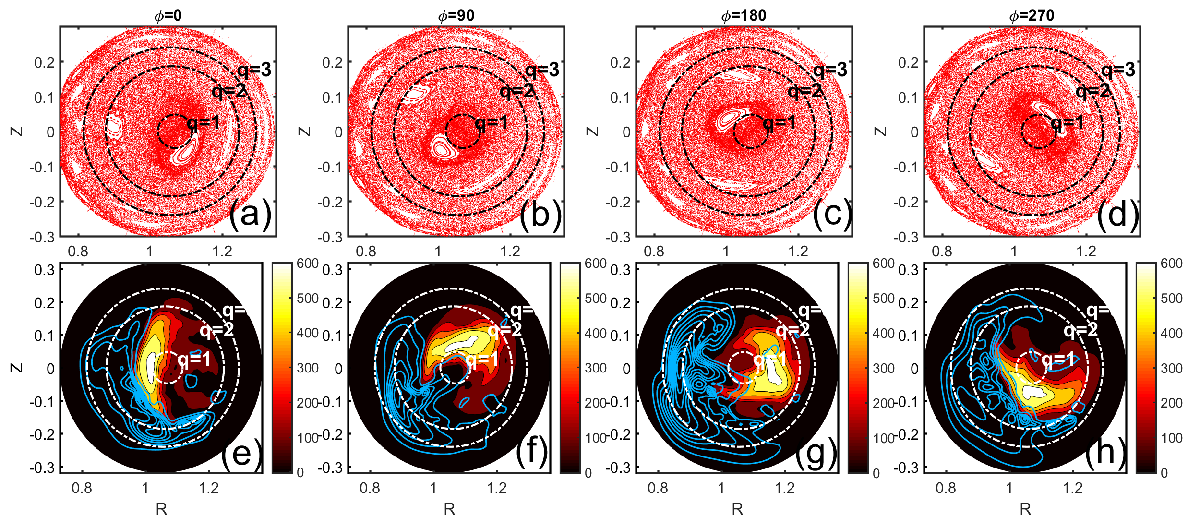}
	\end{center}
	\caption{Upper row: Poincare plots of the magnetic fields including both the equilibrium and the $n=1$ components (red dot) in the poloidal planes at different toroidal locations, where (a)-(d) refer to toroidal angles $\phi=0, 90, 180, 270$ respectively; Lower row: Electron temperature distribution (in unit $eV$, flushed color), and impurity ion distribution (blue line, the sum of all Ar ion charge states) in the poloidal plane at different toroidal locations, where (e)-(h) refer to toroidal angles $\phi=0, 90, 180, 270$ respectively. $q = 1,2,3$ surfaces are denoted as the dashed-line circles. Here $t=1.15ms$.}
	\label{TQ_CB}
\end{figure}

The final stage of TQ begins after $t = 1.05 ms$.
Even during the TQ, only a small fraction of the impurity ion accumulation around the $q = 2$ surface further penetrates near and inside the $q = 1$ surface in the core region (Fig. \ref{TQ_CB}).
The initial impurity penetration is mainly from the isotropic diffusion in absence of initial equilibrium flow. By the time of TQ ($t=1.15ms$), the impurity ion density distribution peaks at poloidal angle $\theta=270$ near the injection location in the poloidal plane at toroidal angle $\phi=0$ (Fig. \ref{TQ_CB}e). Meanwhile, at toroidal angle $\phi=180$ the impurity density distribution peak is located at poloidal angle $\theta=180$ (the HFS) (Fig. \ref{TQ_CB}g). The ratio of angular migration rates of the bulk impurity density distribution near the $q=2$ surface is $\Delta \phi/\Delta \theta = 180/90=2$, as also noted from previous studies \cite{Eidietis2017}.

From the distributions of the impurity ion density in the poloidal planes at different toroidal angles shown in Fig. \ref{TQ_CB} at $t=1.15ms$, one can see that the impurity ion density concentrates within the poloidal angle range $\theta=120-330$ in all poloidal planes. It is from the location around toroidal angle $\phi=90$ that the impurity ion density penetrates into the core region inside $q = 1$ surface to cool down the hot core plasma and gives rise to the cold bubble formation, where the impurity density distribution peak corresponds to the X-point of the $2/1$ mode in the poloidal plane (Fig. \ref{TQ_CB}b and Fig. \ref{TQ_CB}f). Note that the impurity gas tends to enter through only one of the two X-points of the $2/1$ mode, and that is likely the cause for the 1/1 mode structure of the cold bubble.
After initial spreading in the parallel direction toward the HFS,  impurities remain concentrated both poloidally and toroidally as the magnetic field gradient impedes further propagation back toward the LFS. Due to this nozzle effect \cite{Izzo2015} the impurities do not spread much beyond half way in the toroidal direction and $1/4$ of the way around the poloidal plane.

\begin{figure}[ht]
	\begin{center}
		\includegraphics[width=0.45\textwidth,height=0.3\textheight]{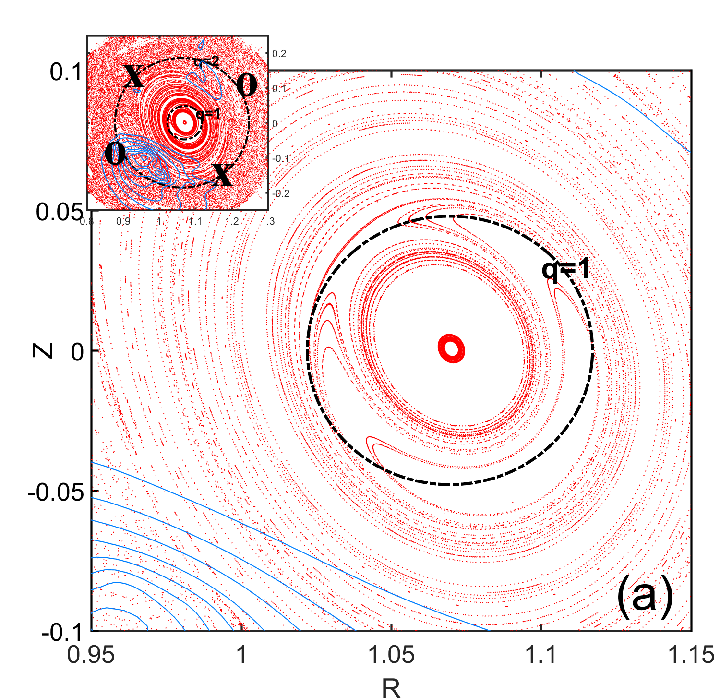}
		\includegraphics[width=0.45\textwidth,height=0.3\textheight]{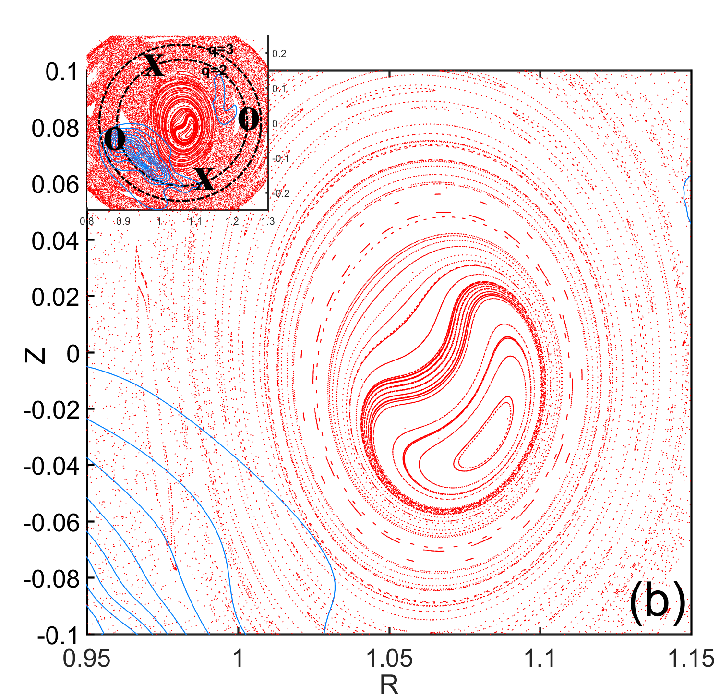}

	\end{center}
	\caption{Poincare plots of the magnetic fields including both the equilibrium and the $n=1$ components (red dot), and impurity ion density distribution (blue line, the sum of all Ar ion charge states) in the $\phi=90$ poloidal planes with different equilibrium $q_0$ case. (a) $q_0=0.95$ and $t=0.75ms$, (b) $q_0=1.1$ and $t=1.0ms$. equilibrium $q = 1,2,3$ surfaces are denoted as the black dashed-line circles.}
	\label{11mode}
\end{figure}

In addition, the Poincare plots of the magnetic fields including both the equilibrium and the $n=1$ components (Fig. \ref{TQ_CB}) provide another phase relationship between the $2/1$ and $1/1$ island with the cold bubble.
A subdominant $1/1$ mode in the central region appears earlier around $t=0.75ms$ before the onset of TQ (Fig. \ref{11mode}a), which is caused by the current contraction in the core region due to the radiation cooling. However, the dominant $1/1$ cold bubble structure in temperature forms only after the impurity penetration into the central region at the beginning of the TQ. Thus the subdominant $1/1$ mode is not the cause of the cold bubble formation. Initially, the O-point of the subdominant $1/1$ mode is aligned with the X-point of the dominant $2/1$ mode in the plane of toroidal angle $\phi=90$ at $t=0.75ms$ (Fig. \ref{11mode}a). Afterward, the poloidal phase of the $1/1$ mode rapidly evolves until the cold bubble formation following the impurity penetration into core region, when the O-point of the $1/1$ mode is locked to the cold plasma region of the cold bubble, and the X-point to the hot spot of plasma expelled from the core region.
Both JET \cite{Lehnen2015a} and J-TEXT \cite{Tong2019} experiments have observed the same phase relationship between the $1/1$ mode and the cold bubble.

\begin{figure}[ht]
  \begin{center}
  \includegraphics[width=1.0\textwidth,height=0.3\textheight]{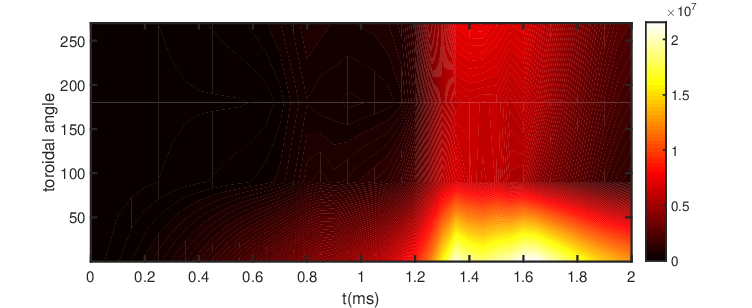}
  \end{center}
  \caption{Toroidal distribution of radiation power (in unit $W$) as a function of time.}
\label{radiation_asymmetry}
\end{figure}

Figs. \ref{TQ_CB}(e)-\ref{TQ_CB}(h) shows that the hot core plasma is expelled from the central region in the poloidal plane leading to the core temperature collapse ($t=1.05ms$), which defines the timing of TQ onset in general. In the $\phi=0$ poloidal plane, the hot core plasma is expelled from exactly the same poloidal angle as the impurity gas cold front. The enhanced interaction between those two contributes to the flash of radiation power and toroidal asymmetry during the TQ shown in Fig. \ref{radiation_asymmetry}, i.e. $t=1.35 ms$. Thus the phase relationship among the $2/1$ mode, the cold bubble, and the impurity ion cold front gives rise to the intrinsic asymmetry in toroidal radiation power distribution by the end of TQ.

\subsection{The effect of the $1/1$ mode}
\begin{figure}[ht]
	\begin{center}
		\includegraphics[width=1.0\textwidth,height=0.3\textheight]{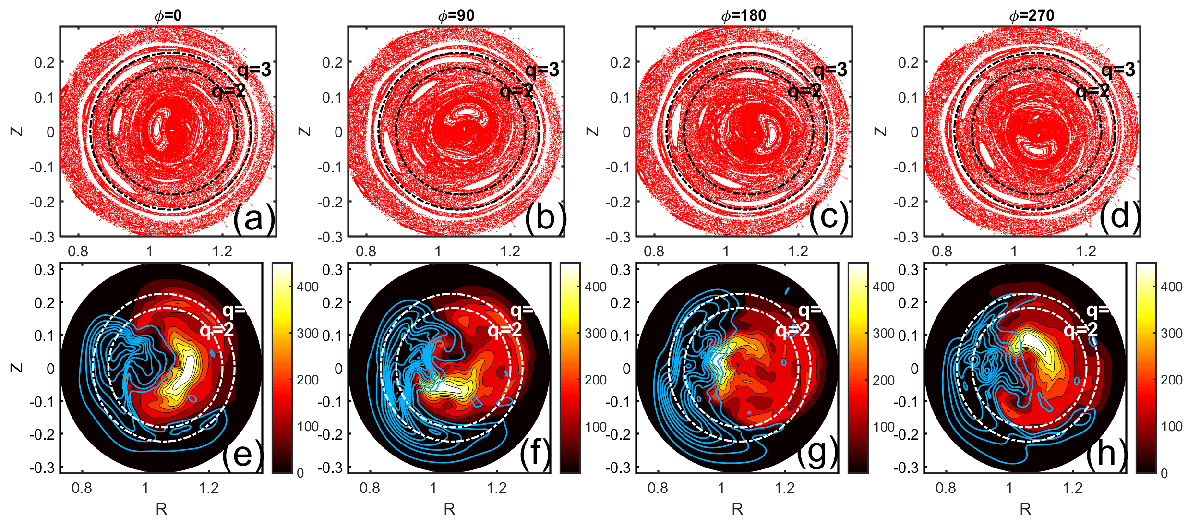}
	\end{center}
	\caption{Upper row: Poincare plots of the magnetic fields including both the equilibrium and the $n=1$ components (red dot) in the poloidal planes at different toroidal locations, where (a)-(d) refer to toroidal angles $\phi=0, 90, 180, 270$ respectively; Lower row: Electron temperature distribution (in unit $eV$, flushed color), and impurity ion distribution (blue line, the sum of all Ar ion charge states) in the poloidal plane at different toroidal locations, where (e)-(h) refer to toroidal angles $\phi=0, 90, 180, 270$ respectively. $q = 2$ and $q=3$ surfaces are denoted as the dashed-line circles. Here is the $q_0=1.1$ case and $t=1.4ms$.}
	\label{q1case}
\end{figure}

In order to clarify the roles of the subdominant $1/1$ mode in the core region, we set up another equilibrium with $q_0>1$, i.e. $q_0 = 1.1$ whereas all other equilibrium properties remain the same. A quasi-interchange mode appears in the central region as shown in Fig. \ref{11mode}(b), which does not involve magnetic reconnection. In this case, the impurity penetrates into the core region and gives rise to the cold bubble formation, however, through the O-point instead of the X-point of the $2/1$ mode (Fig. \ref{q1case}a and Fig. \ref{q1case}e). This difference in the impurity penetration manner may relate to different nature of the subdominant $1/1$ mode.

\section{Discussion and summary}
\label{sec:summary}
In summary, key features of MHD activities often observed in MGI experiments, including the onset of $2/1$ tearing mode and the formation of cold bubble, have been reproduced in our recent NIMROD simulations, and their causal relations have been explored and established in this work.
In these simulations, the plasma thermal energy (radiation power) gradually decreases (increases) after impurity injection. This is followed by the sudden collapse of core electron temperature at the beginning of TQ, and a spike of plasma current near the end of TQ. During the TQ phase, the amplitudes of the $m=2,n=1$ mode and radiation power both reach peak values.

During the pre-TQ stage, magnetic islands are observed to form sequentially after the arrival of impurity ion cold front at the $q = 3$ and the $q = 2$ rational surfaces.
A local current sheet forms at the interface of impurity and plasma upon their direct contact due to radiative cooling.
Impurities rotate with and accumulate in the islands, and are seen to impede further inward penetration.
Our calculations indicate that there is no unstable $2/1$ external mode with or without a wall, mostly due to the low normalized $\beta$ ($\beta_N=0.2639$) and high edge $q$ ($q_a = 3.797$) values of the equilibrium. The dominant $2/1$ mode in our simulation is internal and local to the core region. Thus we do not expect the simulation boundary outside the plasma edge would affect much the simulation results.

The initial impurity inward penetration across flux surface mainly comes from the isotropic diffusion in absence of initial equilibrium flow. On each flux surface, the rapid thermal equilibration along magnetic field lines, i.e. $\nabla_{\parallel} T \simeq 0$, due to the large parallel thermal conductivity $\chi_{\parallel}$ also helps to broaden the parallel spread of the impurity density distribution, i.e. $\nabla_{\parallel} n_z \simeq 0$, as a result of the initial static equilibrium, i.e. $\nabla_{\parallel} p \simeq 0$. In addition, whereas the ratio of impurity angular migration rates between toroidal and poloidal directions is proportional to $q$ \cite{Eidietis2017}, the extent of poloidal spread of the impurity is limited due to the relatively short time scale of thermal quench as shown in, e.g. Fig. \ref{TQ_CB}.
As a consequence, the impurity distribution in the poloidal plane has only enough time to reach one X-point or O-point of the $2/1$ mode by the time right before the cold bubble formation, whereas the impurity spreads at least twice as far in the toroidal angle. Whether the impurity penetration through X-point or O-point of the $2/1$ mode depends on the nature of the subdominant $1/1$ mode in the core region. However, the $1/1$ mode alone is not the cause of the cold bubble formation, which only takes place after the impurity penetration inside the $q=2$ surface following the $2/1$ mode growth.

Despite the establishment of the relations among the $2/1$ mode, the cold bubble, and the impurity penetration in simulations, several key questions on their interaction remain to be addressed. For example, what is the role of the $1/1$ mode appearing inside $q = 1$ surface? How does the nature of the $1/1$ mode affect the particle and energy transport at O-point and X-point? Understanding the dynamic interactions between impurity penetration and magnetic reconnection may provide insights on how to improve the efficiencies of the impurity assimilation process and the disruption mitigation scheme based on the methods of impurity gas injection. We plan to tackle those remaining issues in future work.

\section{Acknowledgments}
We are grateful for the discussions with Prof. C. R. Sovinec, as well as the supports from the NIMROD team and the J-TEXT team.
This work was supported by the National Magnetic Confinement
Fusion Program of China (Grant No. 2019YFE03050004), the
National Natural Science Foundation of China (Grant Nos. 11775221
and 51821005), the Fundamental Research Funds for the Central
Universities at Huazhong University of Science and Technology
(Grant No. 2019kfyXJJS193), and U.S. Department of Energy (Grant
Nos. DE-FG02-86ER53218 and DE-SC0018001). This research used
the computing resources from the Supercomputing Center of
University of Science and Technology of China.

\newpage
\section{Reference}

\providecommand{\newblock}{}


\newpage

\newpage

\newpage

\newpage

\newpage

\newpage

\newpage

\newpage

\newpage

\newpage

\newpage


\end{document}